\documentclass[useAMS,usenatbib]{mn2e}

\usepackage{graphicx}

\usepackage{psfrag}
\usepackage{color}

\def\Ekin{E_{\rm{kin}}}
\def\Egrav{E_{\rm{grav}}} 
\def\alphakin{\alpha_{\mathrm{kin}}} 
\def\solmas{$\mathrm{M_\odot}$~}
\def\solmasp{$\mathrm{M_\odot}$}

\def\tff{$t_{\mathrm{ff}}$}
\def\tffm{t_{\mathrm{ff}}}

\def\be{\begin{equation}}
\def\ee{\end{equation}}

\def\density{g~cm$^{-3}$}

\title[The SFE and its relation to variations in the IMF]
{The star formation efficiency and its relation to variations in the initial mass function}
\author[Clark,  Bonnell \& Klessen]
{Paul C. Clark$^1$ \thanks{E-mail: pcc@ita.uni-heidelberg.de}, 
Ian A. Bonnell$^2$
\& Ralf S. Klessen$^1$
\\
$^1$ Institut f\"ur Theoretische Astrophysik, Universit\"at Heidelberg,
Albert-Ueberle-Stra\ss e. 2, Heidelberg, Germany  \\
$^2$ School of Physics \&
Astronomy, University of St Andrews, North Haugh, St Andrews, Fife, KY16 9SS, UK 
}

\date{\today}

\begin{document}
\maketitle

%
%

\begin{abstract}

We investigate how the dynamical state of a turbulently supported, 1000\,\solmasp,
molecular cloud affects the properties of the cluster it forms, focusing our
discussion on the star formation efficiency (SFE) and the initial mass function
(IMF). A variety of initial energy states are examined in this paper, ranging from
clouds with $|\Egrav| = 0.1\,\Ekin$ to clouds with $|\Egrav| = 10\,\Ekin$ , and for
both isothermal and piece-wise polytropic equations of state (similar to that
suggested by Larson). It is found that arbitrary star formation efficiencies are
possible, with strongly unbound clouds yielding very low star formation
efficiencies. We suggest that the low star formation efficiency in the Maddelena
cloud may be a consequence of the relatively unbound state of its internal
structure. It is also found that competitive accretion results in the observed IMF
when the clouds have initial energy states of $|\Egrav| \ge \Ekin$. We show
that under such conditions the shape of the IMF is independent of time in the
calculations. This demonstrates that the global accretion process can be terminated
at any stage in the cluster's evolution, while still yielding a distribution of
stellar masses that is consistent with the observed IMF. As the clouds become
progressively more unbound, competitive accretion is less important and the
protostellar mass function flattens. These results predict that molecular clouds
should be permeated with a distributed  population of stars that follow a flatter
than Salpeter IMF. 

\end{abstract}

%
%

%
%
%
%
%
%
%
%
%
\section{Introduction}
\label{sec:intro}

\begin{figure*}
\centerline{ 	
	\psfrag{EPSILON}{\textcolor{white}{$\alphakin = 2$}}
	\psfrag{TIME}{\textcolor{white}{Isothermal}}
	\psfrag{EOS}{\textcolor{white}{$t = 0.50 \,\tffm$}}
	\psfrag{LENGTH}{\textcolor{white}{0.25 pc}}
		\includegraphics[width=1.7in,height=1.7in]
			{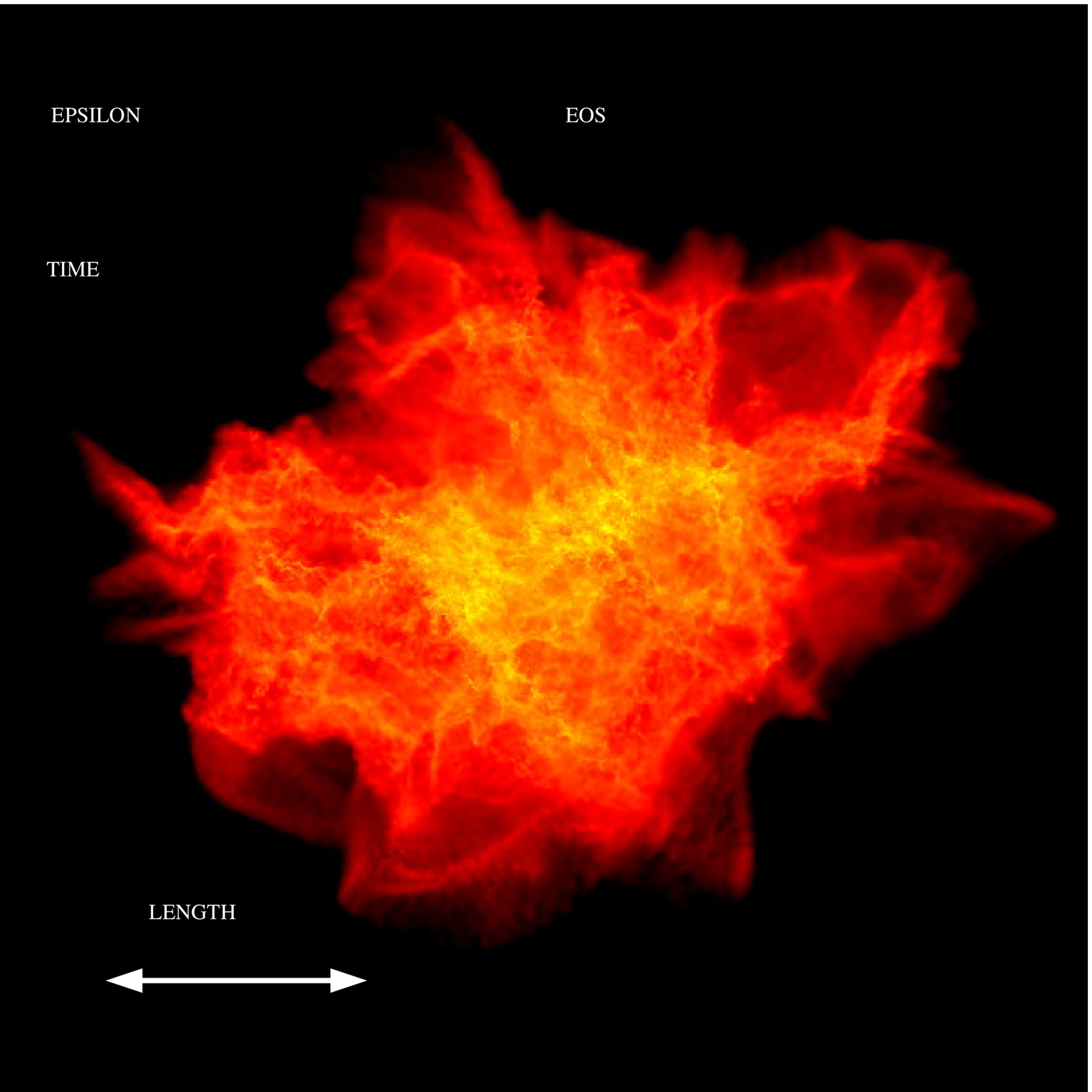}
       \psfrag{EOS}{\textcolor{white}{$t = 1.00 \,\tffm$}}
		\includegraphics[width=1.7in,height=1.7in]
			{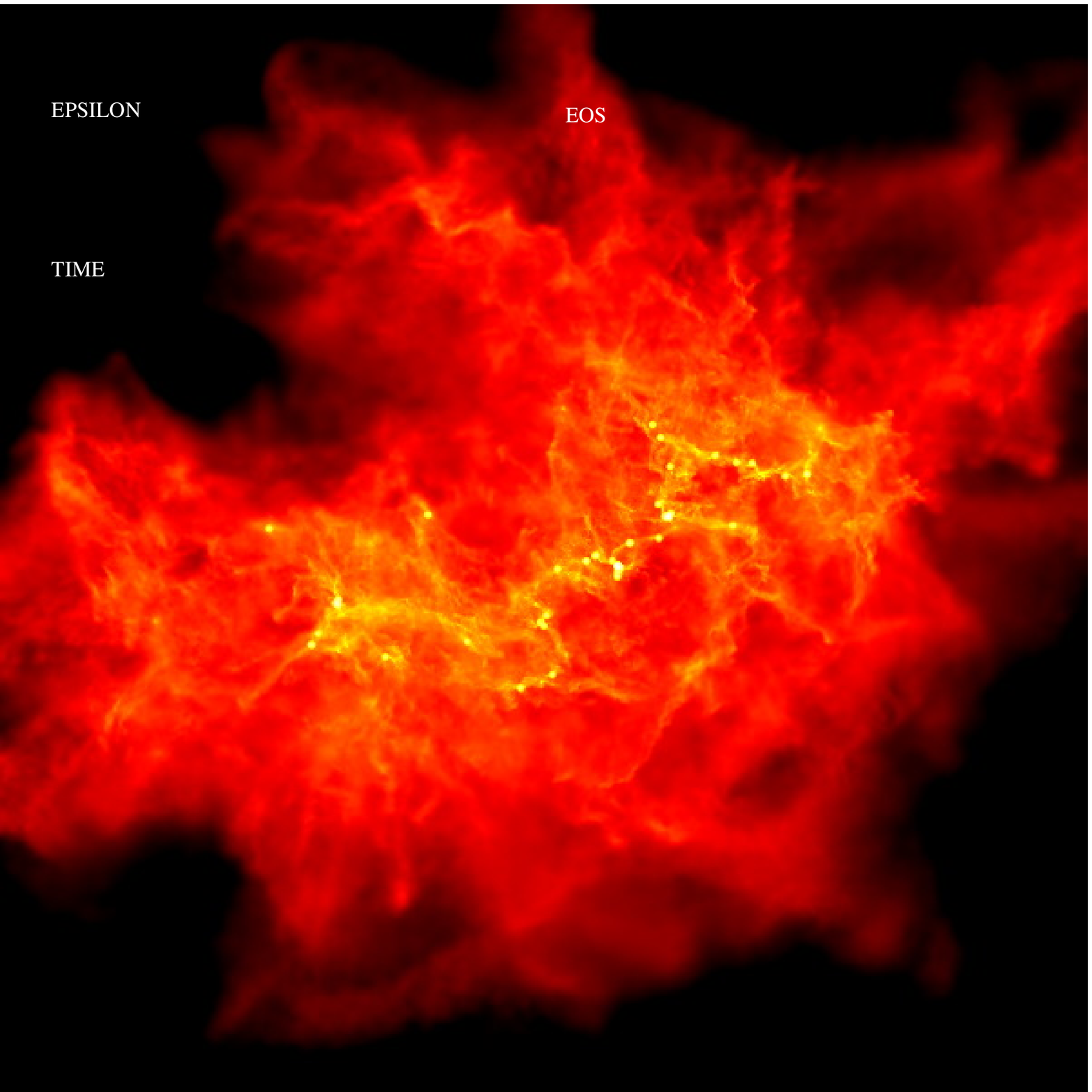}
	\psfrag{EOS}{\textcolor{white}{$t = 1.50 \,\tffm$}}
		\includegraphics[width=1.7in,height=1.7in]
			{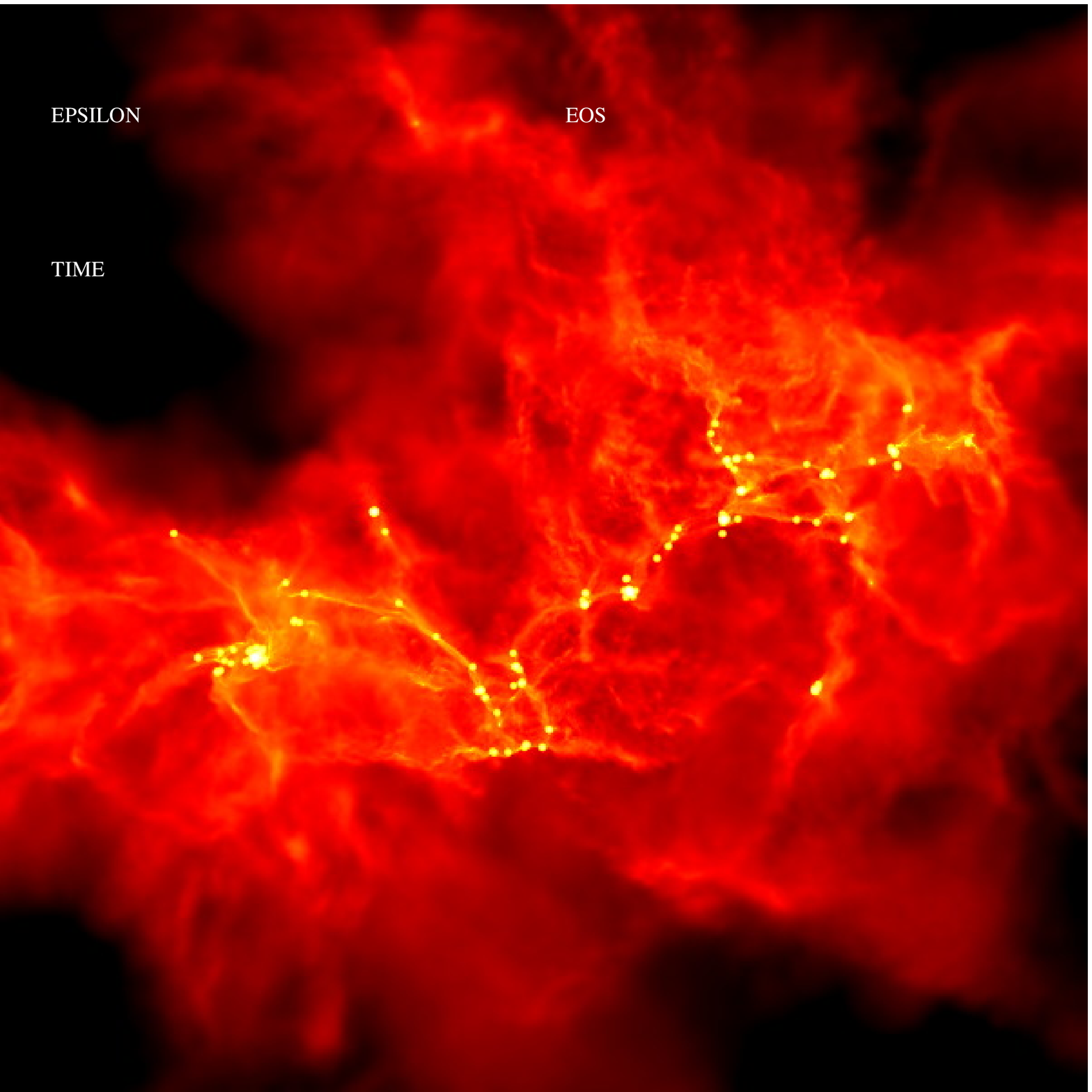}
	\psfrag{EOS}{\textcolor{white}{$t = 2.00 \,\tffm$}}
		\includegraphics[width=1.7in,height=1.7in]
			{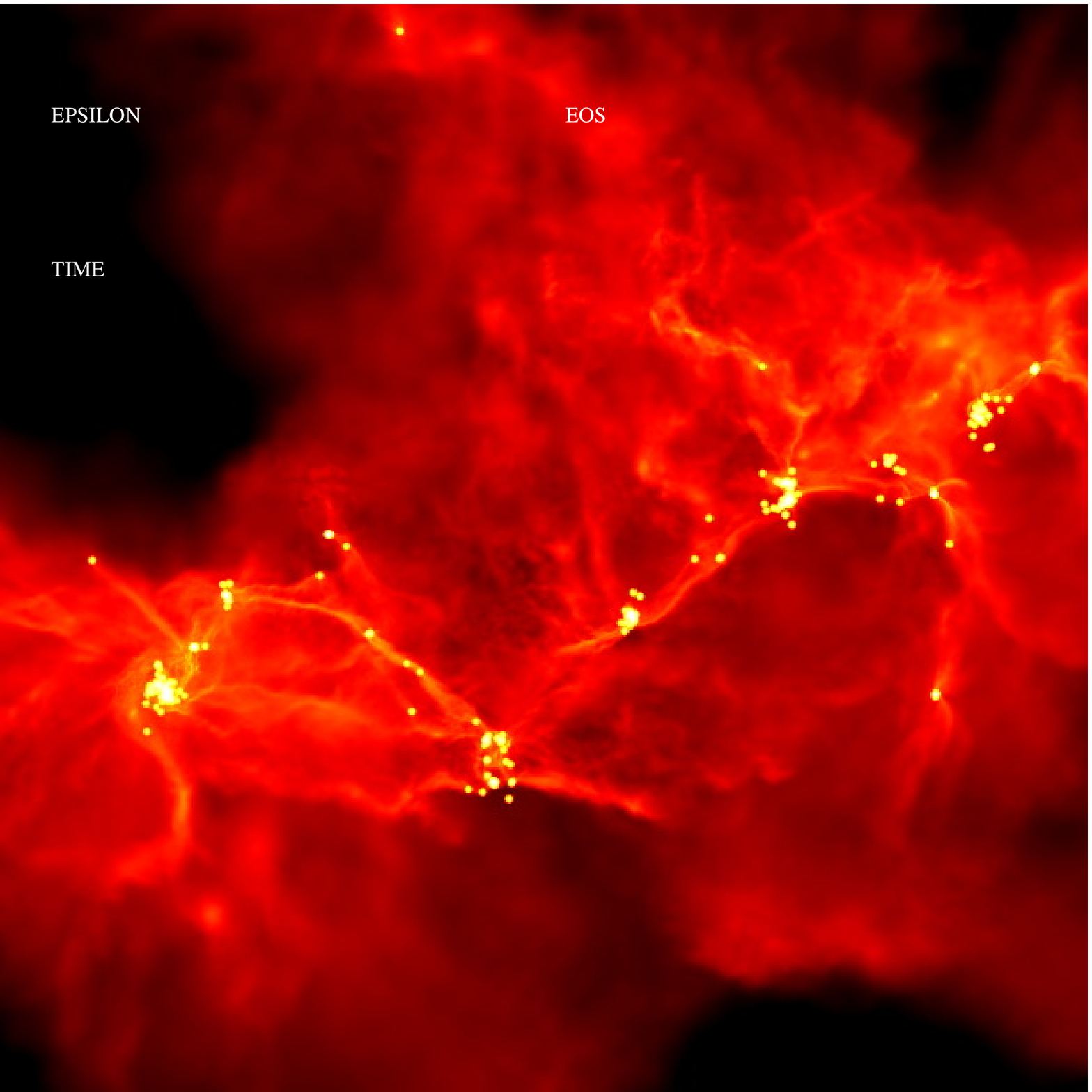}
}
\centerline{ 	
	\psfrag{EPSILON}{\textcolor{white}{$\alphakin = 2$}}
	\psfrag{EOS}{\textcolor{white}{Barotropic EOS}}
	\psfrag{TIME}{\textcolor{white}{$t = 0.50 \tffm$ }}
	\psfrag{LENGTH}{\textcolor{white}{0.5 pc}}
		\includegraphics[width=1.7in,height=1.7in]
			{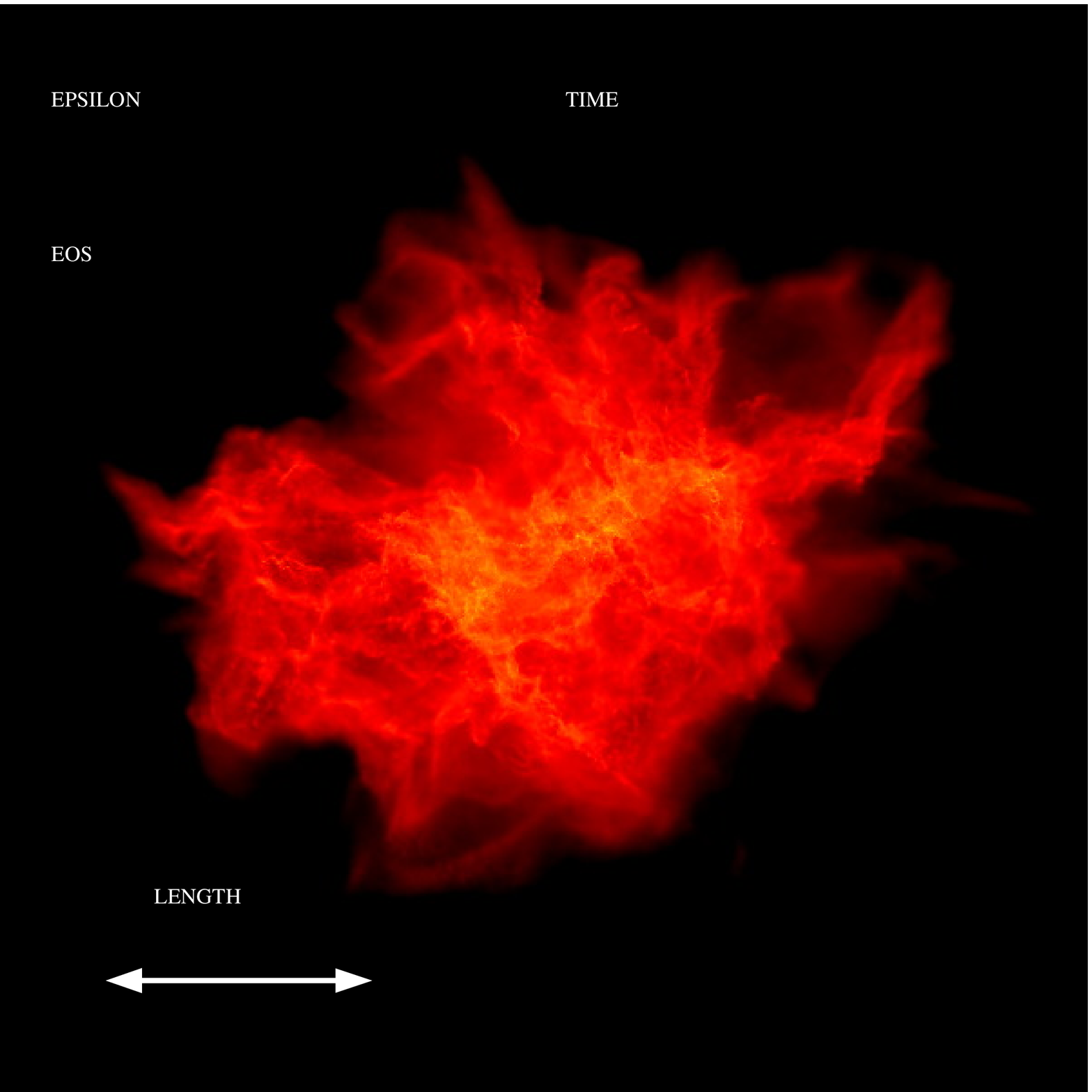}
	\psfrag{TIME}{\textcolor{white}{$t = 1.00 \,\tffm$}}
		\includegraphics[width=1.7in,height=1.7in]
			{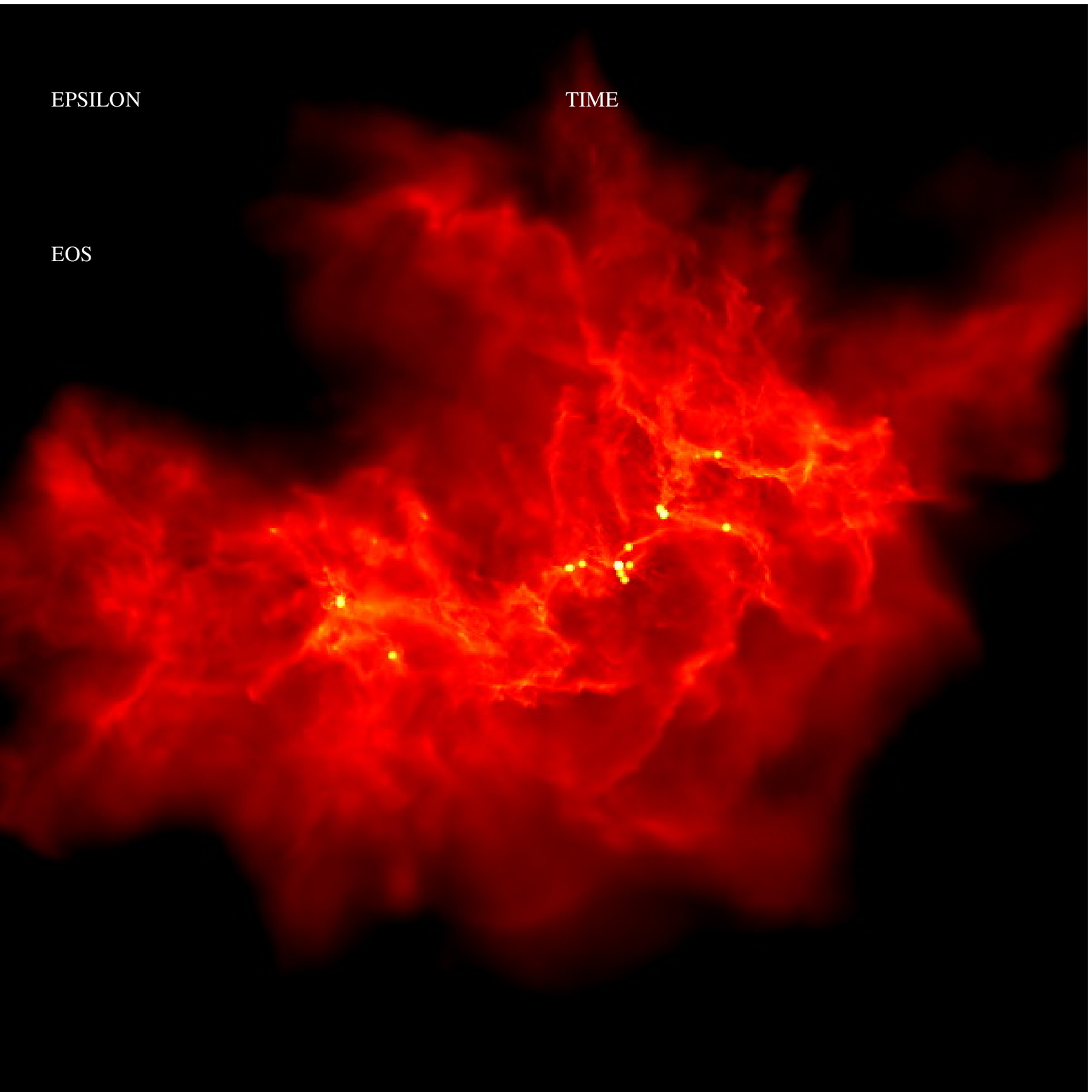}
	\psfrag{TIME}{\textcolor{white}{$t = 1.50 \,\tffm$}}
		\includegraphics[width=1.7in,height=1.7in]
			{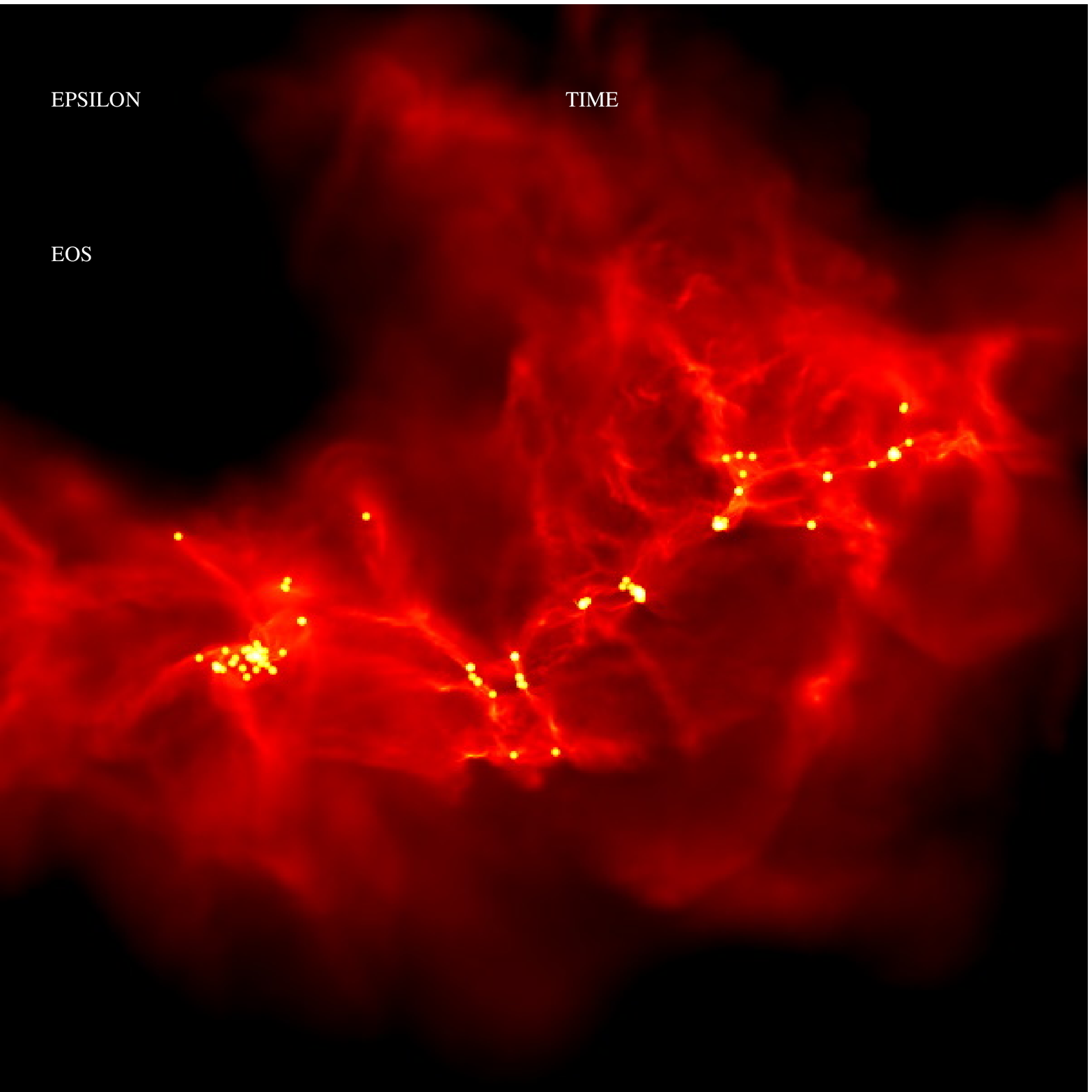}
	\psfrag{TIME}{\textcolor{white}{$t = 2.00 \,\tffm$}}
		\includegraphics[width=1.7in,height=1.7in]
			{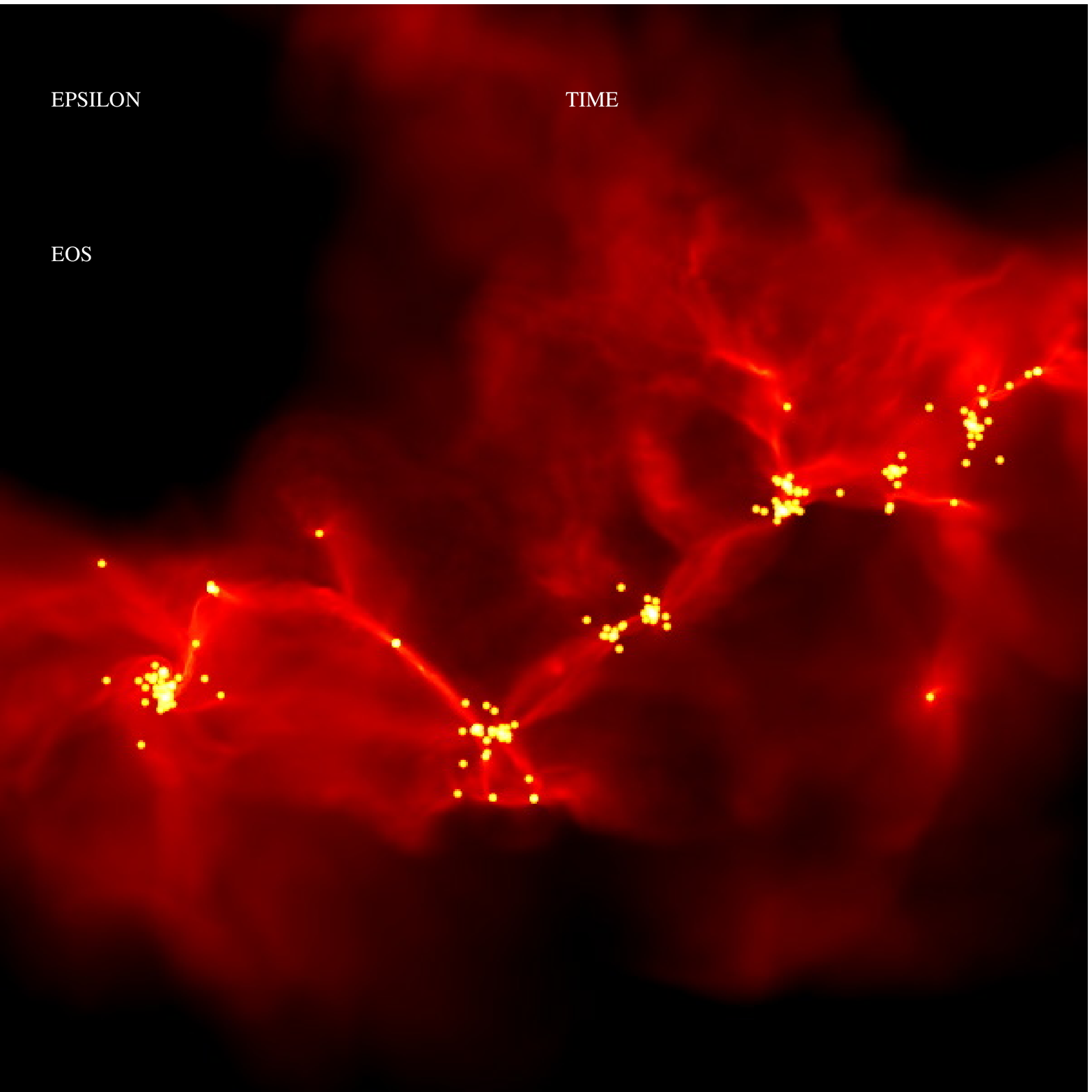}
}

\caption{\label{images} Evolution of the column density distribution in the gas
for 2 of the $\alphakin =\Ekin\,/\,|\Egrav|= 2$ simulations, for an
isothermal (top) and a piece-wise barotropic EOS (bottom), similar to that discussed by Larson
(2005). Both the simulations have the same initial turbulent field. The colour table
in the images is stretched (logarithmically) over 5 orders of magnitude in column
density, ranging from 0.001 to 100 g~cm$^{-2}$.}

\end{figure*}

The internal motions in giant molecular clouds (GMCs) are dominated by supersonic
turbulence (for example,
\citealt{Larson1981,Myers1983,Dameetal1986,Solomonetal1987,HeyerBrunt2004}). These
motions leave their imprint on the cloud's internal structure
\citep{MacLowKlessen2004, ElmegreenScalo2004}, resulting in dense regions, often
referred to as clumps or cores, that are observed to have a wide range of dynamical
states \citep{Ballyetal1987, Williamsetal1994}. The aim of this paper is to explore
how the initial dynamical state of star forming regions can influence both their
final stellar mass functions and their star formation efficiency (SFE).

\citet{ClarkBonnell2004} suggested that stars can form in dynamically unbound
molecular regions, where the internal turbulent motions are in excess of the
region's gravitational energy. Depending on the scale, such regions can be star
forming cores or clouds, and could thus give rise to a wide range of cluster sizes.
For dynamically unbound regions to be capable of producing stars, they need to
contain multiple thermal Jeans masses, or undergo better
than isothermal cooling during the shock compression phase (e.g.
\citealt{Chapmanetal1992, Bhattaletal1998, Larson2005}).  This is simply because
isothermal shock compression (in 1-D) does not significantly increase the number of
Jeans masses in the shock layer \citep*{Elmegreens1978,  Doroshkevich1980,
LubowPringle1993, Clarke1999}. This fact is apparent in the numerical studies of
turbulent molecular clouds, which show that the number of stars is always comparable
to the number of initial Jeans masses in the simulation, despite the huge density
contrast produced by the turbulence or colliding flows
\citep{Klessenetal2000,Gittinsetal2003,ClarkBonnell2005,Larson2007}.

Dynamically unbound star forming regions have three important features. First, star
formation occurs on a similar timescale to the dispersion of the region. Such a
picture is consistent with the rapid star formation model
\citep*{VazquezSemadeni1995,Paredesetal1999,Elmegreen2000, Hartmannetal2001, Pringleetal2001}. 
This process has gained recent support from the studies of
\citet{GloverMacLow2007a,GloverMacLow2007b}, which show that H$_{2}$ formation can
occur on a dynamical time in both turbulent or freely collapsing gas. The second
feature is that unbound clouds can naturally result in lower star formation
efficiencies than their bound counterparts, simply because much of the gas is unable
to rid itself of the excess kinetic energy and achieve a bound state. The
efficiencies can range from 50 percent at protostellar core scales
\citep{ClarkBonnell2004}, to between 5 and 10 percent at the larger scales of a GMC
\citep{Clarketal2005}. Finally, \citet{Clarketal2005} also found that at the GMC
scale, the unbound cloud can produce a series of clusters that reproduce some of the
features found in OB sub-groups.

The aim of this present study is two-fold. First, we wish to explore how the initial
energy balance in the cloud can affect the star formation efficiency, extending the
work of \citet{ClarkBonnell2004}  and \citet{Clarketal2005} (see also
\citealt*{Klessenetal2000}, \citealt*{Heitschetal2001},
\citealt*{VazquezSemadenietal2003}). Secondly, we wish to look at how the mass
function of the protostars is influenced by such an environment. When competitive
accretion (\citealt{Zinnecker1982}; \citealt{Bonnelletal2001a}; 
\citealt{Bonnelletal2001b})
is the dominant process controlling the distribution of stellar masses, it
has been shown that the break in the protostellar mass function, and thus the
characteristic mass, depends on the mean Jeans mass
(\citealt{Klessenetal1998,KlessenBurkert2000,KlessenBurkert2001,Klessen2001,
Jappsenetal2005}; \citealt*{Bonnelletal2006}). For clouds which are out of global
dynamic stability, the mean density (and thus the Jeans mass) are constantly
changing. As such, the ability of competitive accretion to produce an IMF-like
protostellar mass function under these conditions is not clear.

To address these questions we perform a series of numerical simulations which follow
the evolution of 1000\,\solmas clouds, in which the initial ratio of turbulent
kinetic energy to gravitational potential energy is systematically varied. These
clouds, which are similar to those studied by \citet*{BBV2003}, can be thought of as
over-densities in a larger GMC, which still contain residual kinetic energy from the
supersonic flows that created them. Further details of the calculations are given in
Section \ref{simdetails}, and we discuss the star formation efficiency and mass
functions from these clouds in sections \ref{sec:sfes} and \ref{sec:imfs},
respectively. The implications of our results for star formation in GMCs in general
are presented in Section \ref{sec:chat}, and a brief summary of our findings is
given in Section \ref{sec:summary}

\section{The Simulations}
\label{simdetails}
\subsection{Computational Method and Cloud set-up}

The evolution of the gas was modelled using the Smoothed Particle Hydrodynamics
(SPH) method (for a description of this technique, see for example
\citealt{Benz1990}, \citealt{Monaghan1992}, \citealt{Monaghan2005}), and the
simulations we present here were performed on the UK Astrophysical Fluids Facility
(UKAFF), a collection of IBM Power 5 nodes. The code utilises adaptive smoothing
lengths for the gas particles, allowing high density regions to be followed with
increased resolution. Gravity is calculated using a binary tree algorithm, which
also serves to construct SPH neighbour lists.

The clouds in this study have a mass of 1000\,\solmasp, which we model using $2
\times 10^{6}$ SPH particles in an initially uniform density sphere (see below for
the initial density values). With this set-up, the simulations have a mass
resolution of 0.05\,\solmasp, sufficient to capture the fragmentation of the gas to
below the hydrogen burning limit \citep{BateBurkert1997}. The clouds are also given a
turbulent velocity field, which is allowed to decay freely via shocks. The
turbulent velocity fields imposed here are described by a power spectrum of $P(k)
\propto k^{-4}$, which is consistent with the turbulent motions observed in
molecular clouds \citep{Larson1981,MyersGammie1999,HeyerBrunt2004}. In
this paper, we investigate the systematic changes that occur in the properties of
the star formation as the dynamical state of a cloud is altered. We therefore use
the same realisation of the turbulent field for most of the simulations. We do
however perform several simulations with a different turbulent field, to check
that any systematic trends in the star formation properties are not due to any
specific feature of the initial turbulent field. To adjust the dynamical state of
the clouds, we change the initial ratio of kinetic to gravitational potential energy, 
$\alphakin = \Ekin\,/\,|\Egrav|$, to vary from 10 to 0.1.

We adopt two types of equation of state (EOS) in this study: the first assumes the
gas is isothermal and the second is a barotropic EOS, similar to that proposed by
\citet{Larson2005}. In the isothermal gas, we set up the gas to have a Jeans mass of
roughly 1\,\solmasp, such that the cloud has initially 1000 Jeans masses. In these
isothermal runs, the clouds have a temperature of 10\,K, and start at a density of
$1.45 \times 10^{19}$\,\density. For the barotropic EOS, we use

\be
\label{LarsonEOS1}
T  = T_{\rm{c}}\,\frac{\rho}{\rho_{\rm{c}}} ^{-0.25}\:,\,\mathrm{for}\:  \rho\,<\,\rho_{\rm{c}}
\ee

\noindent and then, 

\be
\label{LarsonEOS2}
T  = T_{\rm{c}}\:,\,\mathrm{for}\: \rho\,>\,\rho_{\rm{c}},
\ee

\noindent where $\rho_{\rm{c}} = 5.5 \times 10^{-19}$\,\density ~and 
$T_{\rm{c}} = 7.5$\,K. The values for $T_{\rm{c}}$ and $\rho_{\rm{c}}$, 
ensure that at the point where the barotropic EOS
enters the isothermal regime (that is, at densities higher than $\rho_{\rm{c}}$), the
means Jeans mass is roughly the same as the initial value (roughly 1\,\solmasp) found
in the pure isothermal clouds. The barotropic EOS clouds start at a density of $1.83
\times 10^{-20}$\,\density~and a temperature of 17.6\,K. 

To model the star formation in this study we use `sink' particles, as described by
\citet{Bateetal1995}, which involves replacing dense, self-gravitating, regions of
gas with a point mass. Sink particles are capable of accreting further material and
they can interact with the other particles in the simulation via gravity.  The sinks
are formed in these simulations once a particle and it's neighbours are bound,
collapsing, and within an accretion radius, $h_{acc}$, which is taken here to be 
412\,AU. The sinks are then able to accrete any bound SPH particles which fall inside the
accretion radius. The gravitational interactions between the sink particles and all
other particles in the simulation are also smoothed to $h_{acc}$.  Due to the size
of the sink particles in this study, it should be stressed that their mass functions
presented in this paper should not be thought of as a stellar object IMF. 
Instead, we suggest that they are more representative of a `system' mass 
function \citep*{Kroupaetal1991,Kroupa2002}. 
As such, we will refer to them simply as `sink' mass functions in the remainder of 
this paper.

\subsection{General Evolutionary Properties}
\label{sec:genprops}
For the clouds with $\alphakin \ge 1$, we run the simulations for two
initial free-fall times. In our discussion of these simulations, we will use the
properties at the end of the run. For our weakly supported clouds, which have 
$\alphakin = 0.1$, we only follow the evolution for 1 initial free-fall time. In
these latter simulations, the extreme dynamics and environment make them
computationally very expensive. However, since they also evolve faster, one gains
enough information for our present study.

In Figure \ref{images}, we show column density snapshots from two of the
simulations in this study. Both have the same turbulent velocity seed, and have
$\alphakin = 2$. The evolution of the isothermal cloud is shown in the
upper panels, and the evolution of a cloud with a barotropic EOS is shown in the
bottom panels. The expansion of the cloud is apparent in the images, and one can
see that the clusters move apart as they form, in contrast to the merging process
that has been documented in studies where $\alphakin = 1$ initially
(\citealt{BBV2003}; \citealt*{BVB2004}). One can also see that the two different 
EOSs look very similar, since the sites of star formation are selected by the 
turbulent velocity field.

\section{The star formation efficiency}
\label{sec:sfes}

\begin{figure*}
\centerline{ 	\includegraphics[width=3.in,height=3.in]
{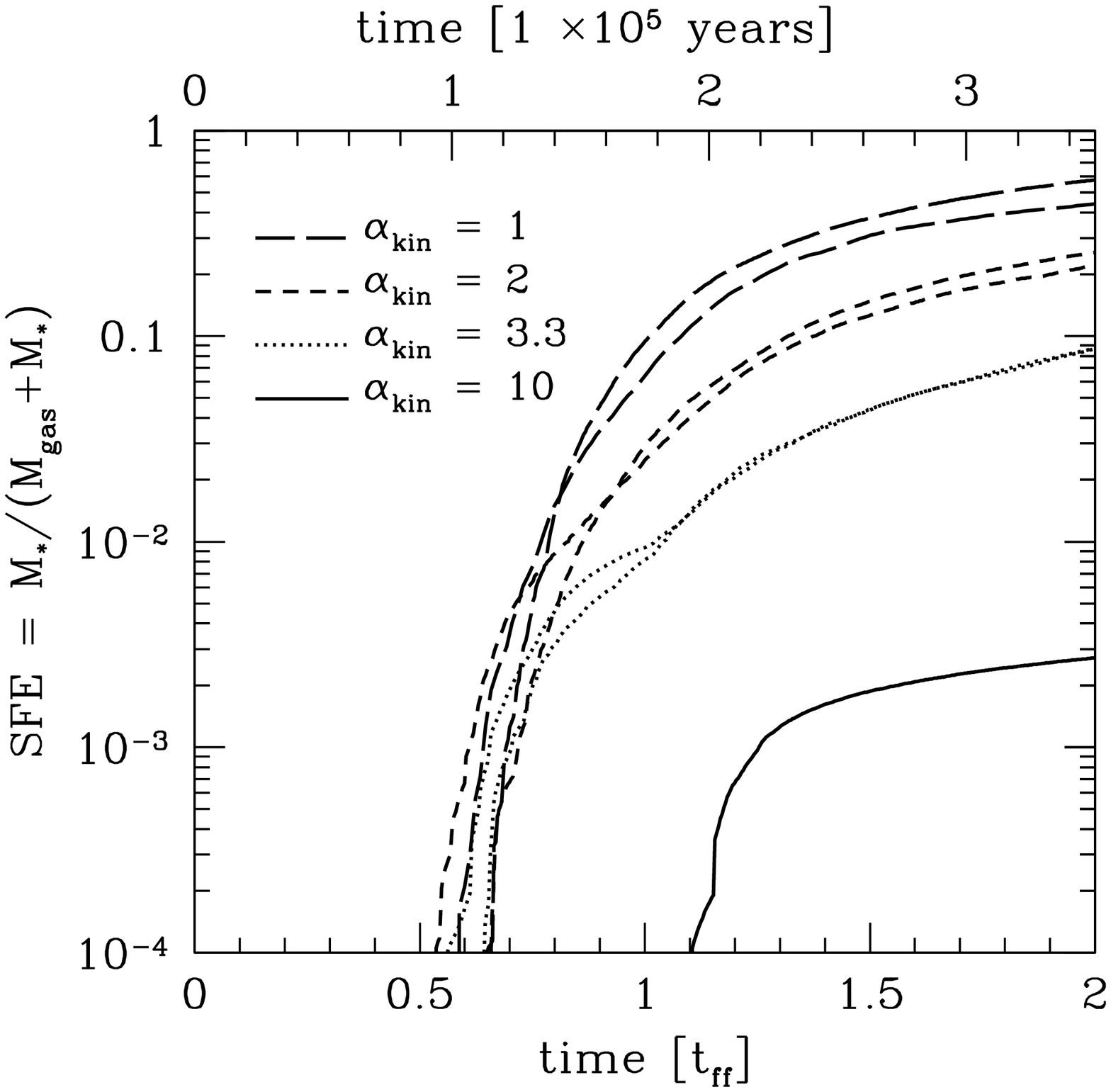}
              \includegraphics[width=3.in,height=3.in]
	{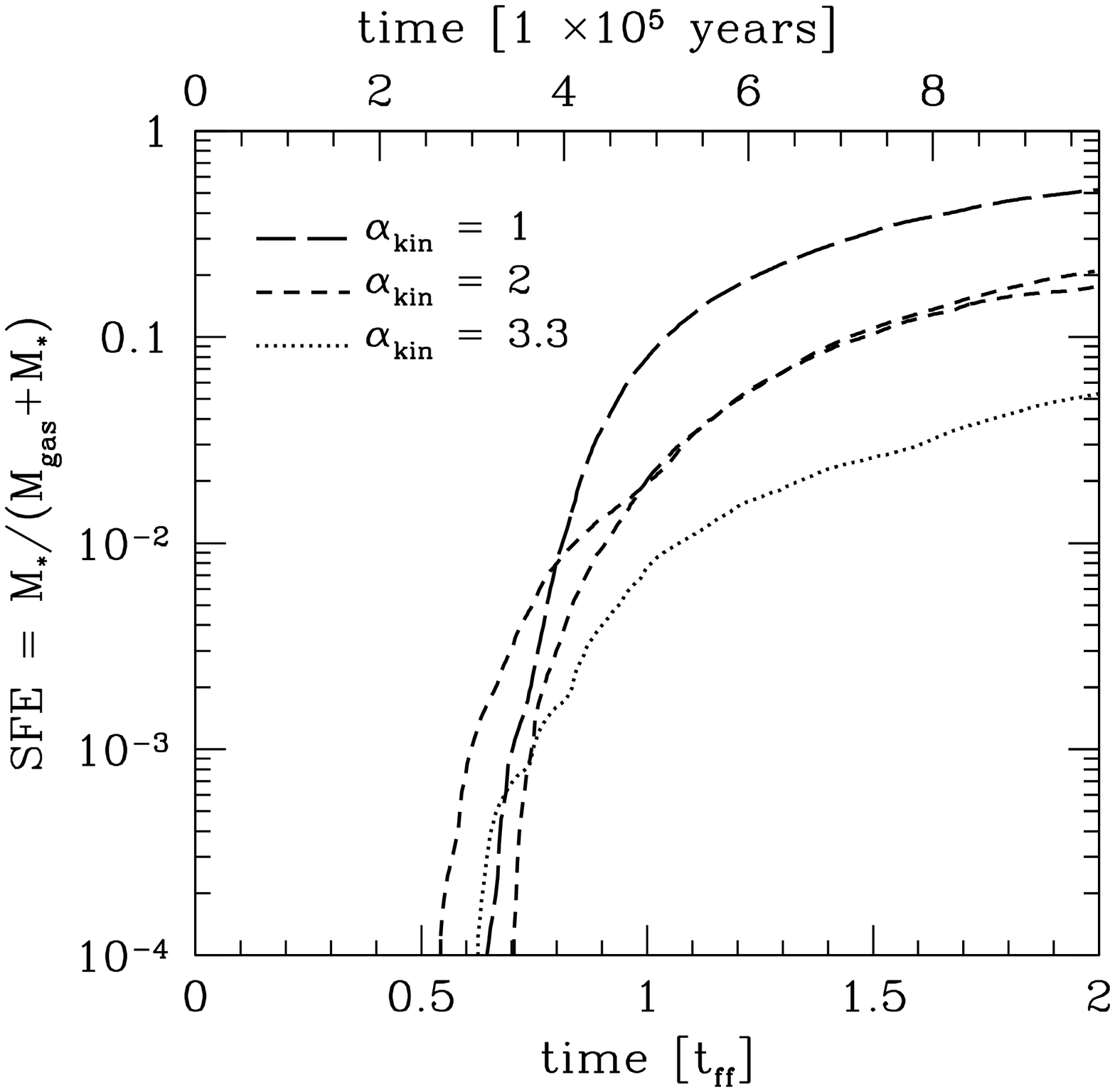}
}
\caption{\label{sfeplots}  The star formation
efficiency, $\mathrm{SFE} = M_{\rm{sinks}}\,/\,(M_{\rm{gas}}\,+\,M_{\rm{sinks}})$, 
as a function of time for the simulations with $\alphakin = \Ekin\,/\,|\Egrav| 
\ge 1$. Two different random
realisations of the initial turbulent velocity field were performed for $\alphakin = $1, 2, 
\& 3.3 in the isothermal case, and for $\alphakin = 2$ in the barotropic case. 
Time is given in both units of a free-fall in the bottom x-axis and in years in the 
top x-axis.}
\end{figure*}

We will first look at how the star formation efficiency (SFE) in a molecular region
can be controlled by its dynamical state. We define the SFE in this study as,
\be
\label{equ:sfe}
\mathrm{SFE} = \frac{M_{\rm{sinks}}}{(M_{\rm{gas}}\,+\,M_{\rm{sinks}})}\:,
\ee
where $M_{sinks}$ and $M_{gas}$ are the total mass in sink particles and gas, 
respectively. In Figure \ref{sfeplots} we plot the SFE
as function of time for all the simulations with $\alphakin = \Ekin\,/\,|\Egrav|
\ge 1$. The isothermal calculations are on the left and the barotropic EOS are on
the right. We plot time in units of both the initial free-fall time of the clouds,
and in years. The free-fall time is $1.75 \times 10^{5}$\,yr in the isothermal runs,
and $4.92 \times 10^{5}$\,yr in the barotropic EOS.

The plots show that a wide range of SFEs are possible, depending on the initial
dynamical state of the cloud, with the SFE decreasing with increasing $\alphakin$.
These calculations also suggest that star formation is able to take
place under a variety of conditions, provided that the region has a sufficient
number of Jeans masses at the outset.While the SFEs in our simulations
are clearly not converged, and without feedback one would not expect them to be,
they do demonstrate a clear trend with the initial level of turbulent
support. One must also consider here that the SFEs given in Figure
\ref{sfeplots} are {\it{upper}} limits to the true value, since sink
particles do not contain any model of the feedback that may be important in
inhibiting further accretion. Also, not all the gas accreted by the sinks would be
expected to find its way into stars, especially for the high-mass objects 
(see \citealt{ZinneckerYorke2007} for further discussions). 

The barotropic and isothermal EOSs give very similar
results, in terms of the SFE evolution as a function of the initial free-fall time.
This suggests that the regions for collapse are selected for fragmentation at
similar times in the evolution, and that these regions proceed directly to collapse
on the corresponding free-fall time. The similarity of both the cloud structure and
cluster distribution between the two different EOS (Figure \ref{images}) also
suggests that the general morphology of the region is controlled primarily by the
turbulent motions flows, via shock compression and dissipation of kinetic energy.
It is also interesting to note just how quickly the SFE drops as the value of
$\alphakin$ increases. Given that most the structure in a GMC is typically thought
to be unbound \citep{Williamsetal1994}, one would expect the majority of
star formation in GMCs to only occur in a few regions, in which the local SFE can be
quite high.

\begin{figure}
\centerline{  \includegraphics[width=3.in,height=3.in]
	{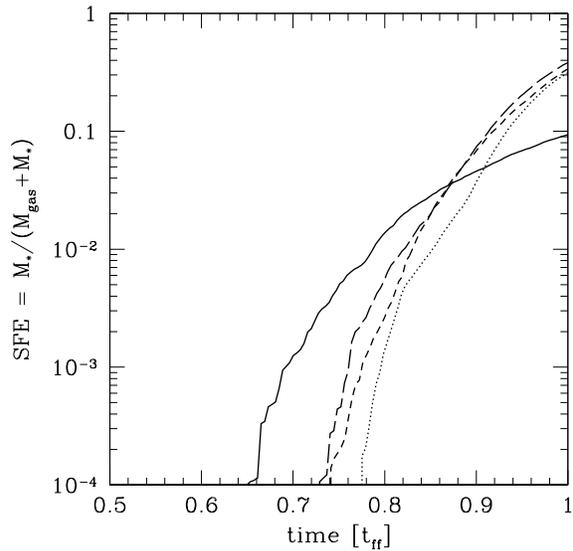}
}
\caption{\label{eps10sfe} The mass in sinks as a function of time
for the clouds with $\alphakin = \Ekin\,/\,|\Egrav| =
0.1$). The long-dashed and short-dashed lines show the $SFE(t)$ for two isothermal
runs which have different initial random realisations of the turbulent velocity field. 
The dotted line traces the SFE($t$) for the simulation with
the barotropic EOS. For comparison, we plot  the SFE($t$) from the
isothermal $\alphakin = 1$ simulation (solid line), which shares the same seed as the
$\alphakin = 0.1$ denoted by the long-dashed line.  }
\end{figure}

The weakly supported clouds, naturally, have a much higher rate of star formation.
We plot in Figure \ref{eps10sfe} the SFE as function of time (free-fall units) for
the clouds with $\alphakin = 0.1$, and for comparison, we also show the SFE
curve from the isothermal $\alphakin = 1$ calculation. Clearly, these weakly
supported clouds evolve much faster than their more strongly supported counterparts, 
reaching 3-4 times the efficiency in only half the time.

\section{The sink particle mass functions}
\label{sec:imfs}

\begin{figure*}
\includegraphics[width=6.in,height=4.1812in]{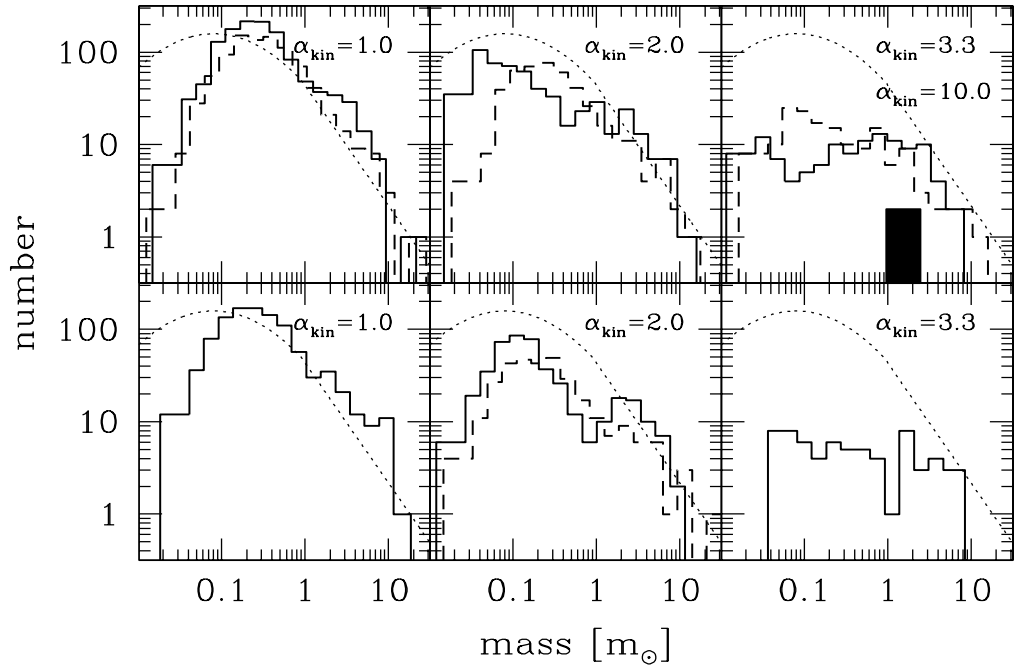}

\caption{\label{imfUB} 
The mass functions of the sink particles for the simulations with $\alphakin =
\Ekin\,/\,|\Egrav| \ge 1$, given after a time $t = 2\,t_{\rm{ff}}$. The simulations with an 
isothermal EOS are shown in the
upper panels, while those with the barotropic EOS are shown in the bottom panels.
The long-dashed distributions show the results from a different realisation of the
initial turbulent seed. The isothermal $\alphakin = 0.1$ distribution (filled
region) is shown alongside the $\alphakin = 0.3$ results. The dotted distribution
in each panel is the functional form of the IMF for the local field stars given by
Chabrier (2003). Note that the mass resolution in all cases is
0.05\,\solmasp, and those sinks with masses below this value are still accreting their
neighbours.}
\end{figure*}

As mentioned in Section \ref{sec:intro} above, the effect of the competitive
accretion process on the IMF in dynamically unbound simulations is still unclear, 
and needs to be examined. In Figure \ref{imfUB} we plot the sink particle mass
function for all the simulations that have $\alphakin = \Ekin\,/\,|\Egrav| \ge 1$. 
The isothermal clouds are shown on the left and the barotropic EOS clouds on the right.
Some properties of the sink particles at the end of simulations are given in Table 
\ref{tab:results}.

The panels on the left-hand side show the sink mass functions for the calculations with
$\alphakin = 1$. These have a similar initial set-up to those performed by
\citet{BBV2003, BVB2004, Bonnelletal2006}, where it was demonstrated that for a mean initial
Jeans of roughly 1\,\solmasp, the resulting sink mass function is similar to the field-star IMF
(for the driven turbulent case, see also \citealt{Klessen2001}). 
Our simulations yield a similar result. Although the mean
initial Jeans mass is much higher in the barotropic EOS simulation, the final result is very
similar to that from the isothermal calculation. As has been discussed by others
\citep{Larson2005, Jappsenetal2005, Bonnelletal2006}, this is simply because the 
Jeans mass at the critical density, $\rho_{\mathrm{c}}$, where fragmentation can occur,
is similar to that in the isothermal case.

One can see from Figure \ref{imfUB}, that there is a distinct trend in the sink mass
functions as the value of $\alphakin$ is increased: the mass functions become
flatter as the clouds become progressively more unbound. A similar behaviour is seen
in calculations of driven turbulence with decreasing driving scale
\citep{Klessen2001}. This feature is also seen in the barotropic EOS calculations,
suggesting that the trend in the mass function is not simply due to the changing
Jeans mass in the unbound clouds. One can see this trend in the properties of
the sinks, given in table \ref{tab:results}. The quantity
$f_{\rm{M\,>\,1\,M}_{\odot}}$ denotes the fraction of the total mass in sink
particles that is made up of sinks with masses 1\,\solmas or greater.
\citet{Meyeretal2000} performed a similar analysis on a number of local embedded
clusters (using the number of stars, rather than the mass). One can see here that
the fraction of the mass function comprising sinks of 1\,\solmas or more is clearly
rising as $\alphakin$ is increased, indicating an overall flattening of the mass
functon.

There are two effects which will play an
important role in shaping the sink mass function in the unbound clouds: the
timescale of interactions, and the degree of fragmentation.
First, we consider the timescale on which newly formed objects interact with their
siblings. In the $\alphakin = 1$ simulations, the clouds are characterised by a
particular density. For the isothermal clouds, this is the initial mean cloud
density, while in the barotropic EOS clouds, this is density, $\rho_{\rm{c}}$, at which
the cooling stops. When fragmentation occurs at this density, the timescale for
neighbouring fragments to interact is the same as the timescale for each fragment to
collapse to form a sink particle. Naturally, this is just the free-fall time
associated with the density where the fragmentation sets in. In contrast, the
background density in the unbound simulations is constantly decreasing. The regions
which become bound, and so manage to form sink particles, decouple from the
surrounding material, which is evolving on progressively longer timescales. As a
result, the star-forming regions which form in the unbound clouds are able to use up
their gas reservoir without as much influence from their neighbours. The unbound
clouds hence move towards a more {\it{isolated}} mode of star formation, whereby
individual regions of star formation evolve independently.

The second effect which shapes the mass function is the degree of fragmentation.
From Figure \ref{imfUB}, one can see that the more unbound simulations still form fairly
high-mass sinks. What makes the mass function progressively flatter is the lack of
lower-mass objects, since the clouds with higher levels of turbulent energy form
systematically fewer bound objects. This occurs because many regions take several
sound crossing times to become bound, during which time the internal energy has
erased much of the structure. Naturally, this is more pronounced in the barotropic
EOS calculations, since the gas heats up as it becomes less dense. The lack of
fragmentation within the star forming regions reduces the role of competitive
accretion.

A similar effect is seen in the driven turbulent simulations of \citet{Klessen2001}.
When the turbulent driving scale is small, the stars tend to
form in relative isolation. Since they have little gravitational interaction with
their siblings, the resulting mass functions are flat. In contrast, large scale
driving leads to a highly clustered environment, in which competitive accretion is
important. Consequently, these latter simulations produce mass functions which are more
similar to the IMF.

\begin{table}
\label{tab:results}
\caption{We give here the properties of the sink particles after 2 initial free
fall times have elapsed, for all the simulations with $\alphakin \ge 1$. 
For comparison, we also show the results from one of the simulations with
$\alphakin = 0.1$, with the results taken after 1 initial free-fall time (see
Section \ref{sec:genprops} for details). The equation of state (EOS) is shown in column
2, with `iso' denoting isothermal, and `lar' denoting a piece-wise barotropic EOS 
(see Section \ref{simdetails} and Equations \ref{LarsonEOS1} and \ref{LarsonEOS2}) that
is similar to that suggested by Larson (2005). The seed column shows which of the two
random realisations of the initial turbulent velocity field was used for that 
calculation. $N_{*}$, $M_{*,\,\rm{total}}$ and $\bar{M}_{*}$ are the total number and 
mass in sink particles
and mean sink particle mass, respectively, with $\bar{M_{*}} = M_{*,\,\rm{total}}\,/\,N_{*}$.
Finally, $f_{\rm{M\,>\,1\,M}_{\odot}}$ denotes the fraction of
the total mass in sinks which is made up of sinks with masses of 1 \solmas or greater. 
}
\begin{center}
\begin{tabular}{r|c|c|r|r|r|c}
\hline
$\alphakin$ &EOS & Seed & $N_{*}$  & $M_{*,\,\rm{total}}$  & $\bar{M_{*}}$  & $f_{\,\rm{M\,>\,1\,M}_{\odot}}$\\
\hline \hline
1.0            &iso   & 1  & 1236     & 574.85          & 0.47           &  0.54 \\
2.0            &iso   & 1  & 555      & 255.50          & 0.46           &  0.69 \\
3.3            &iso   & 1  & 120      & 87.54           & 0.73           &  0.70 \\
10.0           &iso   & 1  & 2        & 2.72            & 1.36           &  1.00 \\

1.0            &lar   & 1  & 1029     & 518.72          & 0.50           &  0.57 \\
2.0            &lar   & 1  & 442      & 208.71          & 0.47           &  0.75 \\
3.3            &lar   & 1  & 57       & 53.01           & 0.93           &  0.84 \\

1.0            &iso   & 2  & 899      & 574.85          & 0.47           &  0.54 \\
2.0            &iso   & 2  & 440      & 255.50          & 0.46           &  0.69 \\
3.3            &iso   & 2  & 157      & 87.54           & 0.73           &  0.70 \\
		 
2.0		 &lar   & 2  & 314      & 119.06          & 0.56           &  0.68 \\

0.1		 &iso   & 1  & 1073     & 339.00          & 0.32           &  0.28 \\

\hline
\end{tabular}
\end{center}
\end{table}

\begin{figure}
\centerline{  \includegraphics[width=3.in,height=3.in]
	{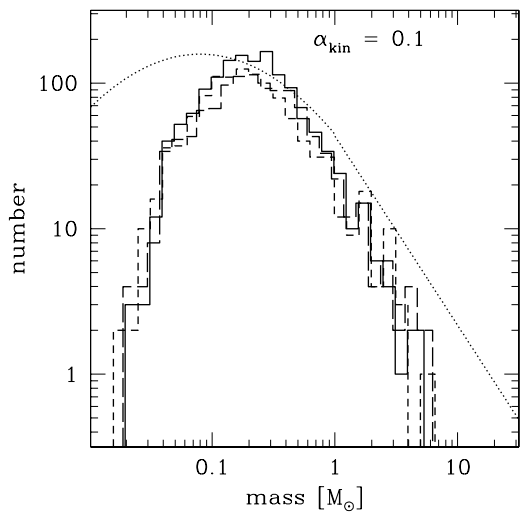}
}

\caption{\label{imfVB} The sink mass functions for the weakly supported systems
(those with $\alphakin\,=\,\Ekin\,/\,|\Egrav|\,=\,0.1$), given after a time $t\,=\,
t_{\rm{ff}}$. The solid and dashed distributions show the mass function for the
isothermal runs, for two different seeds, and the dotted distribution is for the
simulation with the barotropic EOS.The dotted distribution in each panel is the
functional form of the IMF for the local field stars given by \citet{Chabrier2003}.
Note that the mass resolution in all cases is 0.05\,\solmasp, and those sinks with
masses below this value are still accreting their neighbours.}

\end{figure}

After considering the unbound clouds, one may expect that the clouds with $\alphakin
= 10$ to result in mass functions which are very steep. Figure \ref{imfVB}, shows
that this is  not the case.  In these clouds, where the collapse proceeds in a
quasi-homologous fashion,  the sinks interact strongly with their neighbours, which
leads to a mass spectrum that is broadly similar to the IMF. Indeed, it has
been shown \citep{Zinnecker1982, Bonnelletal2001b} that the mass function should
never get infinitely steep under these conditions, with an asymptotic limit  for the
slope reaching $-2.5$. Again we see that the isothermal and barotropic EOSs give
very similar results.

\begin{figure}
\centerline{  \includegraphics[width=3.in,height=3.in]
	{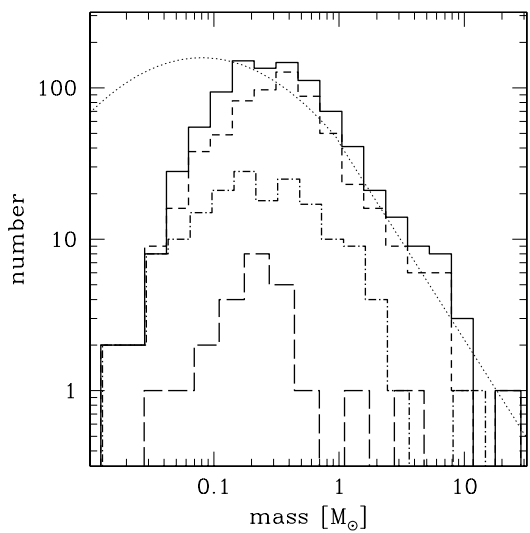}
}

\caption{\label{fig:imfevol} The mass function of sink particles, at four
different instances,  for isothermal simulation with velocity seed number 2 and
$\alphakin = 1$.  The times for which the mass functions are shown are 0.75
(long-dashed),  1 (dot-dashed), 1.5 (short-dashed) and 2 (solid) initial free-fall
times. Note that while the mass function at 2\,\tff is the same as plotted in Figure 
\ref{imfUB} (top left panel, dashed distribution), the binning here is slightly
larger, simply to make the plot easier to read. Again, the dotted line denotes the
Chabrier (2003) IMF.} 
\end{figure}

Lastly, we note that when competitive accretion occurs in clouds with 
$\alphakin \le 1$, the mass function of the accreting objects always 
has roughly the same shape. This in demonstrated in Figure \ref{fig:imfevol}, 
which shows the evolution of the mass function for one of the isothermal runs with 
$\alphakin = 1$. This is an important feature of the mass function in these
cloud, since it means that the star formation need not be terminated at some
special time, but can indeed be terminated at any point in the evolution and will 
still give the correct IMF.

One can see why the shape of the mass function remains constant in the
$\alphakin \le 1$ calculations by considering the time-scales for
accretion and fragmentation in these calculations. As discussed by
\citet{Bonnelletal2001a}, the accretion time-scale in such clouds is comparable
to the free-fall time. Further, the crossing time, which dictates 
the decay time-scale of the turbulence in the gas \citep{MacLowetal1998}, 
is related to the free-fall time in
these clouds by,
\be
\frac{\tffm}{t_{\rm{cr}}} = \left( \frac{3\,\pi}{32\,G\,\rho} \right)^{1/2}\: \frac{2\,r}{v_{\rm{rms}}}
= \frac{\pi}{\sqrt{10}}\,\alphakin^{1/2} \:,
\ee 

where $r$ is the radius of the cloud, $v_{\rm{rms}}$ is the root-mean-square of the
turbulent velocities, and $\rho$ is the mean density in the cloud. Naturally, this
is also the average time-scale on which regions of gas will lose their turbulent
support and start to collapse, and as such can be thought of as the fragmentation
time.  Since the time-scales for the accretion and fragmentation are similar
throughout the cloud's evolution, the rate at which new sinks form is then similar
to rate at which mass is added to the pre-existing sinks via accretion. Thus, both
the  Salpeter section of the mass function and the flatter section at lower masses
grow together, which maintains the shape of the mass distribution.

In the contrasting picture to competitive accretion, in which the stellar masses are
set purely by the fragmentation of the molecular cloud, it is not clear that the
emerging IMF would be time independent. \citet*{Clarketal2007} pointed out that any
stellar mass function that results from the collapse of a bound distribution of
cores will change with time, provided that the time-scale for the cores to collapse
is a function of their mass. While this may not be a problem for the
\citet{Salpeter1955} section of the core mass function, the time-scale issue may
change the mapping between the observed cores and final stars at lower masses (for
example, see \citealt{Andreetal2007}).

\section{Discussion}
\label{sec:chat}

The results of this paper show that arbitrarily low star formation efficiencies are
possible with just slight variations in the initial kinematic state of the cloud or
star forming region. As determinations of the total cloud masses and kinetic energy
content are uncertain, such variations are easily within our current understanding
of molecular clouds. For example, an initial $\alphakin=0.3$ corresponds to a
mass three times less than what is required to be bound or equivalently an internal
velocity dispersion 1.8 times higher.

In order to fully understand the significance of these results, we need to put them
in a broader context of cloud formation and evolution. Recent work has shown that a
dynamical formation  mechanism for molecular clouds is likely due to some larger
scale compression \citep{Paredesetal1999, Hartmannetal2001, Bonnelletal2006b,
Dobbsetal2006, VazquezSemadenetal2006}. Shocks induced by spiral arms, or supernovae
are two such possibilities. In these cases, there is no reason to assume that the
clouds are gravitationally bound on the largest scales. Instead, as local regions
become {\em close} to being gravitationally bound due to the compression
\citep{Klessenetal2005,BallesterosParedes2006}, then star formation is initiated on
this local scale (Bonnell et al 2006). The star formation is likely to be
inefficient except in those regions which are strongly gravitationally dominated,
resulting in a low overall star formation efficiency.

Although the above scenario is dynamical in nature, it can easily be mistaken as
being {\sl slow} in terms of the star formation rate per free-fall time
\citep{KrumholzTan2007}. The primary
reason for this is that an observer is unlikely to see only the final stages of the
process but is equally likely to observe any intermediate state in the evolution.
Thus what we need to compute is the time-averaged star formation efficiency for a
given cloud. For example, given the end results presented in Figure \ref{sfeplots},
the time averaged star formation efficiency per free-fall time (as defined from the 
initial density) for the cases of $\alphakin\,=~1,~0.5,~0.3$ are 
SFE$\,/\,t_{\rm{ff}}\,\approx\,0.18$, 0.07, and 0.02,
respectively. These are significantly lower than just the final star formation
efficiency and could be taken to imply that star formation in these simulations was
slow. Furthermore, we need  also consider the formation timescale of the cloud which
effectively doubles the cloud lifetime \citep{Paredesetal1999}  and thus halves the
above estimates of the star formation efficiency per free-fall time to $0.09, 0.035$
and 0.01.

The results of this paper may also provide an explanation for the relative lack of star
formation in the Madellena GMC. We show here that star
formation becomes less efficient and more distributed as the clouds become more
unbound. Similarly, \citet{Williamsetal1994} have demonstrated that the structure in
the Maddelena cloud is significantly unbound, and such structures have estimated
masses in the range from around 2 to 1000\,\solmasp. We hence suggest that the highly
unbound and dynamic environment in the Maddelena cloud may be responsible for the
very low level of star formation.

We also predict that dynamically unbound regions should produce more isolated stars
that follow a flatter IMF than those in the bound clustered environments. Given the
unbound state of the majority of the structure in, for example, the Orion A cloud
\citep{Ballyetal1987} or the Rosette molecular cloud \citep{Williamsetal1994},
we would predict that a low level of star formation should permeate most molecular
clouds, providing that the dense regions have a sufficient number of Jeans masses.
Current observations suggest that this is indeed the case. The 2MASS and
{\em{Spitzer}} surveys have revealed distributed populations of stars in a number of
nearby star-forming regions \citep{Carpenter2000, Allenetal2007}, and the data
suggest a smooth transition between `distributed' and `clustered' star formation
\citep{Allenetal2007}. From the wide variety in dynamical states for the
sub-structure in molecular clouds, our results would also suggest a smooth
transition between heavily clustered environments and regions of relatively isolated
star formation.

One potential problem with the model we present here is that the resulting
field-star IMF in the Galactic disc would become contaminated by the flatter IMFs
from the unbound regions. If most star  formation occurs in the more massive bound
clusters, then competitive accretion is able to produce the observed IMF. In
contrast, if the majority of stars instead form in loose associations, in which
gravity only dominates at the scale of individual stars or systems, then our results
would suggest an integrated IMF that is shallower than the observed distribution.
However, the star formation efficiency is also important. Although the more unbound
clouds produce flatter IMFs, they do so with a lower efficiency, and thus the
overall effect of the distributed population is not clear. Estimates for the
distributed population in several nearby star forming regions suggest that these
objects comprise around 20 to 30 percent of the host cloud's star formation
\citep{Allenetal2007}, which will also include any stars that have left dense
clusters or groups. This implies that the contamination problem is not too severe.
We also note that \citet{Ladas2003} estimate that the majority of stars do form in
clusters.

Lastly, we issue a word of caution when interpreting the mass functions presented in
this study. Sink particles only store the {\em potential} mass reservoir for the
protostar, or protostellar system, and do not allow material to escape once it has
been accreted. In reality stars will have winds and outflows, which are expected to
lower the SFE of the protostellar core. However for low-mass stars, this effect is
likely to be small \citep{WuchterlKlessen2001}. The formation of higher-mass stars
may be less efficient due to the increased radiation pressure and presence of
ionising winds \citep*{YorkeSonnhalter2002, Krumholzetal2005b}. Therefore, if the 
{\em internal} SFE of a sink particle is a decreasing function of increasing mass,
then the final stellar IMF would indeed be shallower than those presented in this
study. To recover the observed Salpeter slope, the sink particle mass  function
would then need to be somewhat steeper. The sinks used in this study also have
accretion radii that are larger than typical binary separations
\citep{DuquennoyMayor1991}. As such, they are more representative of multiple
systems, for systems smaller than the accretion radius $r_{\rm{acc}}$ (as we point
out in section \ref{simdetails}). One would therefore expect the mass function to
steepen somewhat for higher resolution.

\section{Summary}
\label{sec:summary}

This paper examines the properties of the star formation in a series of 1000\,\solmas
clouds, in which the initial level of turbulent support is altered, with
$\alphakin =  \Ekin\,/\,|\Egrav|$ varying from 0.1 to 10. We perform these
calculations for two different EOSs, first assuming isothermality and then assuming
a barotropic EOS, the latter of which is similar to that suggested by Larson (2005).
The starting density in the barotropic EOS clouds is over an order of magnitude
lower than that in the isothermal clouds.

We find that a wide range of star formation efficiencies is possible. In the
simulations for which $\alphakin \ge 1$, the star formation efficiencies range
from up to 60 percent, to as low as around 0.3 percent, after a time of 2 initial
free-fall times. Since these calculations do not contain any model for the feedback from
young stars, these SFEs are upper limits.
The isothermal and barotropic equations of state yield similar
results. We suggest that the very low level of star formation in the Maddelena cloud
may be a result of the highly unbound structure that has been observed in this
object \citet{Williamsetal1994}. In contrast, the simulations with $\alphakin =
0.1$ are able to reach a star formation efficiency of around 35 percent, in only 1
free-fall time.

The mass functions of the `sink' particles formed in the simulations are also found
to be related to each cloud's initial dynamical state. In the clouds which start
with $\alphakin \ge 1$, the mass functions of the sinks becomes progressively
flatter with increasing initial energy in the turbulence. For the clouds with
$\alphakin = 1$ to 0.1, the sink particle mass function are broadly similar to the
stellar IMF. However we caution the reader that the sink particles used in this
study are larger than the typical binary system, and so are best interpreted as a
system mass, which would be expected to be flatter than a true stellar mass
function. All the features seen in the mass functions in this study are parallelled
in both the EOSs that are employed.

We also point out that for $\alphakin \le 1$, the general shape of the mass
function is constant in time, including position of the turnover. This is an
important feature of the mass function: regardless of when competitive accretion is 
interrupted by gas expulsion, the resulting mass function will be consistent with
the observed distribution. As such, it does not suffer from the `time-scale'
problem, discussed in detail by \citet{Clarketal2007}, which can arise when trying
to map a pre-stellar core mass function to a star or system mass function.

We conclude that if competitive accretion is to play an important role in shaping
the stellar IMF, then the majority of star formation needs to occur in bound,
collapsing regions, at the scale of the cluster formation. However, the GMCs that
host these cluster forming regions need not be globally bound entities. Within the
GMC's unbound sub-structure, we predict that there can be a distributed mode of star
formation, provided the gas in these structures has a sufficient number of Jeans
masses. These stars are expected to exist in relative isolation, with a trend
towards progressively more isolation in more unbound or dynamic regions. We predict
that the IMF of this distributed population is flatter than that found in either the
clustered environment or the local field star population.

\section{Acknowledgements}

We thank the anonymous referee for a close reading of the original manuscript which
helped to clarify a number of issues. P.C.C. acknowledges support by the Deutsche 
Forschungsgemeinschaft (DFG) under grant KL 1358/5 and via the 
Sonderforschungsbereich (SFB) SFB 439, Galaxien im fr\"uhen Universum.
We would like to thank Chris Rudge and Richard West at the UK Astrophysical Fluid
Facility (UKAFF) for their tireless assistance and enthusiasm during the completion
of this work.

%
%
 
\bibliographystyle{mn2e}
\bibliography{/home/spyros/0/pcc/bibfile/pccbib}

\begin{thebibliography}{}

\bibitem[\protect\citeauthoryear{{Allen}, {Megeath}, {Gutermuth}, {Myers},
  {Wolk}, {Adams}, {Muzerolle}, {Young} \& {Pipher}}{{Allen}
  et~al.}{2007}]{Allenetal2007}
{Allen} L.,  {Megeath} S.~T.,  {Gutermuth} R.,  {Myers} P.~C.,  {Wolk} S.,
  {Adams} F.~C.,  {Muzerolle} J.,  {Young} E.,    {Pipher} J.~L.,  2007, in
  {Reipurth} B.,  {Jewitt} D.,   {Keil} K.,  eds, Protostars and Planets V {The
  Structure and Evolution of Young Stellar Clusters}.
pp 361--376

\bibitem[\protect\citeauthoryear{{Andr{\'e}}, {Belloche} \&
  {Peretto}}{{Andr{\'e}} et~al.}{2007}]{Andreetal2007}
{Andr{\'e}} P.,  {Belloche} A.,    {Peretto} N.,  2007, in prep

\bibitem[\protect\citeauthoryear{{Ballesteros-Paredes}}{{Ballesteros-Paredes}}%
{2006}]{BallesterosParedes2006}
{Ballesteros-Paredes} J.,  2006, \mnras, 372, 443

\bibitem[\protect\citeauthoryear{{Ballesteros-Paredes}, {Hartmann} \& {V{\'
  a}zquez-Semadeni}}{{Ballesteros-Paredes} et~al.}{1999}]{Paredesetal1999}
{Ballesteros-Paredes} J.,  {Hartmann} L.,    {V{\' a}zquez-Semadeni} E.,  1999,
  \apj, 527, 285

\bibitem[\protect\citeauthoryear{{Bally}, {Stark}, {Wilson} \&
  {Langer}}{{Bally} et~al.}{1987}]{Ballyetal1987}
{Bally} J.,  {Stark} A.~A.,  {Wilson} R.~W.,    {Langer} W.~D.,  1987, \apjl,
  312, L45

\bibitem[\protect\citeauthoryear{{Bate}, {Bonnell} \& {Price}}{{Bate}
  et~al.}{1995}]{Bateetal1995}
{Bate} M.~R.,  {Bonnell} I.~A.,    {Price} N.~M.,  1995, \mnras, 277, 362

\bibitem[\protect\citeauthoryear{{Bate} \& {Burkert}}{{Bate} \&
  {Burkert}}{1997}]{BateBurkert1997}
{Bate} M.~R.,  {Burkert} A.,  1997, \mnras, 288, 1060

\bibitem[\protect\citeauthoryear{{Benz}}{{Benz}}{1990}]{Benz1990}
{Benz} W.,  1990, in Numerical Modelling of Nonlinear Stellar Pulsations
  Problems and Prospects {Smooth Particle Hydrodynamics - a Review}.
p.~269

\bibitem[\protect\citeauthoryear{{Bhattal}, {Francis}, {Watkins} \&
  {Whitworth}}{{Bhattal} et~al.}{1998}]{Bhattaletal1998}
{Bhattal} A.~S.,  {Francis} N.,  {Watkins} S.~J.,    {Whitworth} A.~P.,  1998,
  \mnras, 297, 435

\bibitem[\protect\citeauthoryear{{Bonnell}, {Bate}, {Clarke} \&
  {Pringle}}{{Bonnell} et~al.}{2001a}]{Bonnelletal2001a}
{Bonnell} I.~A.,  {Bate} M.~R.,  {Clarke} C.~J.,    {Pringle} J.~E.,  2001a,
  \mnras, 323, 785

\bibitem[\protect\citeauthoryear{{Bonnell}, {Bate} \& {Vine}}{{Bonnell}
  et~al.}{2003}]{BBV2003}
{Bonnell} I.~A.,  {Bate} M.~R.,    {Vine} S.~G.,  2003, \mnras, 343, 413

\bibitem[\protect\citeauthoryear{{Bonnell}, {Clarke} \& {Bate}}{{Bonnell}
  et~al.}{2006}]{Bonnelletal2006}
{Bonnell} I.~A.,  {Clarke} C.~J.,    {Bate} M.~R.,  2006, \mnras, 368, 1296

\bibitem[\protect\citeauthoryear{{Bonnell}, {Clarke}, {Bate} \&
  {Pringle}}{{Bonnell} et~al.}{2001b}]{Bonnelletal2001b}
{Bonnell} I.~A.,  {Clarke} C.~J.,  {Bate} M.~R.,    {Pringle} J.~E.,  2001b,
  \mnras, 324, 573

\bibitem[\protect\citeauthoryear{{Bonnell}, {Dobbs}, {Robitaille} \&
  {Pringle}}{{Bonnell} et~al.}{2006}]{Bonnelletal2006b}
{Bonnell} I.~A.,  {Dobbs} C.~L.,  {Robitaille} T.~P.,    {Pringle} J.~E.,
  2006, \mnras, 365, 37

\bibitem[\protect\citeauthoryear{{Bonnell}, {Vine} \& {Bate}}{{Bonnell}
  et~al.}{2004}]{BVB2004}
{Bonnell} I.~A.,  {Vine} S.~G.,    {Bate} M.~R.,  2004, ArXiv Astrophysics
  e-prints

\bibitem[\protect\citeauthoryear{{Carpenter}}{{Carpenter}}{2000}]{Carpenter200%
0}
{Carpenter} J.~M.,  2000, \aj, 120, 3139

\bibitem[\protect\citeauthoryear{{Chabrier}}{{Chabrier}}{2003}]{Chabrier2003}
{Chabrier} G.,  2003, \pasp, 115, 763

\bibitem[\protect\citeauthoryear{{Chapman}, {Pongracic}, {Disney}, {Nelson},
  {Turner} \& {Whitworth}}{{Chapman} et~al.}{1992}]{Chapmanetal1992}
{Chapman} S.,  {Pongracic} H.,  {Disney} M.,  {Nelson} A.,  {Turner} J.,
  {Whitworth} A.,  1992, \nat, 359, 207

\bibitem[\protect\citeauthoryear{{Clark} \& {Bonnell}}{{Clark} \&
  {Bonnell}}{2004}]{ClarkBonnell2004}
{Clark} P.~C.,  {Bonnell} I.~A.,  2004, \mnras, 347, L36

\bibitem[\protect\citeauthoryear{{Clark} \& {Bonnell}}{{Clark} \&
  {Bonnell}}{2005}]{ClarkBonnell2005}
{Clark} P.~C.,  {Bonnell} I.~A.,  2005, \mnras, 361, 2

\bibitem[\protect\citeauthoryear{{Clark}, {Bonnell}, {Zinnecker} \&
  {Bate}}{{Clark} et~al.}{2005}]{Clarketal2005}
{Clark} P.~C.,  {Bonnell} I.~A.,  {Zinnecker} H.,    {Bate} M.~R.,  2005,
  \mnras, 359, 809

\bibitem[\protect\citeauthoryear{{Clark}, {Klessen} \& {Bonnell}}{{Clark}
  et~al.}{2007}]{Clarketal2007}
{Clark} P.~C.,  {Klessen} R.~S.,    {Bonnell} I.~A.,  2007, \mnras, 379, 57

\bibitem[\protect\citeauthoryear{{Clarke}}{{Clarke}}{1999}]{Clarke1999}
{Clarke} C.~J.,  1999, \mnras, 307, 328

\bibitem[\protect\citeauthoryear{{Dame}, {Elmegreen}, {Cohen} \&
  {Thaddeus}}{{Dame} et~al.}{1986}]{Dameetal1986}
{Dame} T.~M.,  {Elmegreen} B.~G.,  {Cohen} R.~S.,    {Thaddeus} P.,  1986,
  \apj, 305, 892

\bibitem[\protect\citeauthoryear{{Dobbs}, {Bonnell} \& {Pringle}}{{Dobbs}
  et~al.}{2006}]{Dobbsetal2006}
{Dobbs} C.~L.,  {Bonnell} I.~A.,    {Pringle} J.~E.,  2006, \mnras, 371, 1663

\bibitem[\protect\citeauthoryear{{Doroshkevich}}{{Doroshkevich}}{1980}]{Dorosh%
kevich1980}
{Doroshkevich} A.~G.,  1980, \azh, 57, 259

\bibitem[\protect\citeauthoryear{{Duquennoy} \& {Mayor}}{{Duquennoy} \&
  {Mayor}}{1991}]{DuquennoyMayor1991}
{Duquennoy} A.,  {Mayor} M.,  1991, \aap, 248, 485

\bibitem[\protect\citeauthoryear{{Elmegreen}}{{Elmegreen}}{2000}]{Elmegreen200%
0}
{Elmegreen} B.~G.,  2000, \apj, 530, 277

\bibitem[\protect\citeauthoryear{{Elmegreen} \& {Elmegreen}}{{Elmegreen} \&
  {Elmegreen}}{1978}]{Elmegreens1978}
{Elmegreen} B.~G.,  {Elmegreen} D.~M.,  1978, \apj, 220, 1051

\bibitem[\protect\citeauthoryear{{Elmegreen} \& {Scalo}}{{Elmegreen} \&
  {Scalo}}{2004}]{ElmegreenScalo2004}
{Elmegreen} B.~G.,  {Scalo} J.,  2004, \araa, 42, 211

\bibitem[\protect\citeauthoryear{{Gittins}, {Clarke} \& {Bate}}{{Gittins}
  et~al.}{2003}]{Gittinsetal2003}
{Gittins} D.~M.,  {Clarke} C.~J.,    {Bate} M.~R.,  2003, \mnras, 340, 841

\bibitem[\protect\citeauthoryear{{Glover} \& {Mac Low}}{{Glover} \& {Mac
  Low}}{2007a}]{GloverMacLow2007a}
{Glover} S.~C.~O.,  {Mac Low} M.-M.,  2007a, \apjs, 169, 239

\bibitem[\protect\citeauthoryear{{Glover} \& {Mac Low}}{{Glover} \& {Mac
  Low}}{2007b}]{GloverMacLow2007b}
{Glover} S.~C.~O.,  {Mac Low} M.-M.,  2007b, \apj, 659, 1317

\bibitem[\protect\citeauthoryear{{Hartmann}, {Ballesteros-Paredes} \&
  {Bergin}}{{Hartmann} et~al.}{2001}]{Hartmannetal2001}
{Hartmann} L.,  {Ballesteros-Paredes} J.,    {Bergin} E.~A.,  2001, \apj, 562,
  852

\bibitem[\protect\citeauthoryear{{Heitsch}, {Mac Low} \& {Klessen}}{{Heitsch}
  et~al.}{2001}]{Heitschetal2001}
{Heitsch} F.,  {Mac Low} M.-M.,    {Klessen} R.~S.,  2001, \apj, 547, 280

\bibitem[\protect\citeauthoryear{{Heyer} \& {Brunt}}{{Heyer} \&
  {Brunt}}{2004}]{HeyerBrunt2004}
{Heyer} M.~H.,  {Brunt} C.~M.,  2004, \apjl, 615, L45

\bibitem[\protect\citeauthoryear{{Jappsen}, {Klessen}, {Larson}, {Li} \& {Mac
  Low}}{{Jappsen} et~al.}{2005}]{Jappsenetal2005}
{Jappsen} A.-K.,  {Klessen} R.~S.,  {Larson} R.~B.,  {Li} Y.,    {Mac Low}
  M.-M.,  2005, \aap, 435, 611

\bibitem[\protect\citeauthoryear{{Klessen}}{{Klessen}}{2001}]{Klessen2001}
{Klessen} R.~S.,  2001, \apj, 556, 837

\bibitem[\protect\citeauthoryear{{Klessen}, {Ballesteros-Paredes},
  {V{\'a}zquez-Semadeni} \& {Dur{\'a}n-Rojas}}{{Klessen}
  et~al.}{2005}]{Klessenetal2005}
{Klessen} R.~S.,  {Ballesteros-Paredes} J.,  {V{\'a}zquez-Semadeni} E.,
  {Dur{\'a}n-Rojas} C.,  2005, \apj, 620, 786

\bibitem[\protect\citeauthoryear{{Klessen} \& {Burkert}}{{Klessen} \&
  {Burkert}}{2000}]{KlessenBurkert2000}
{Klessen} R.~S.,  {Burkert} A.,  2000, \apjs, 128, 287

\bibitem[\protect\citeauthoryear{{Klessen} \& {Burkert}}{{Klessen} \&
  {Burkert}}{2001}]{KlessenBurkert2001}
{Klessen} R.~S.,  {Burkert} A.,  2001, \apj, 549, 386

\bibitem[\protect\citeauthoryear{{Klessen}, {Burkert} \& {Bate}}{{Klessen}
  et~al.}{1998}]{Klessenetal1998}
{Klessen} R.~S.,  {Burkert} A.,    {Bate} M.~R.,  1998, \apjl, 501, L205

\bibitem[\protect\citeauthoryear{{Klessen}, {Heitsch} \& {Mac Low}}{{Klessen}
  et~al.}{2000}]{Klessenetal2000}
{Klessen} R.~S.,  {Heitsch} F.,    {Mac Low} M.,  2000, \apj, 535, 887

\bibitem[\protect\citeauthoryear{{Kroupa}}{{Kroupa}}{2002}]{Kroupa2002}
{Kroupa} P.,  2002, Science, 295, 82

\bibitem[\protect\citeauthoryear{{Kroupa}, {Gilmore} \& {Tout}}{{Kroupa}
  et~al.}{1991}]{Kroupaetal1991}
{Kroupa} P.,  {Gilmore} G.,    {Tout} C.~A.,  1991, \mnras, 251, 293

\bibitem[\protect\citeauthoryear{{Krumholz}, {McKee} \& {Klein}}{{Krumholz}
  et~al.}{2005}]{Krumholzetal2005b}
{Krumholz} M.~R.,  {McKee} C.~F.,    {Klein} R.~I.,  2005, \apjl, 618, L33

\bibitem[\protect\citeauthoryear{{Krumholz} \& {Tan}}{{Krumholz} \&
  {Tan}}{2007}]{KrumholzTan2007}
{Krumholz} M.~R.,  {Tan} J.~C.,  2007, \apj, 654, 304

\bibitem[\protect\citeauthoryear{{Lada} \& {Lada}}{{Lada} \&
  {Lada}}{2003}]{Ladas2003}
{Lada} C.~J.,  {Lada} E.~A.,  2003, \araa, 41, 57

\bibitem[\protect\citeauthoryear{{Larson}}{{Larson}}{1981}]{Larson1981}
{Larson} R.~B.,  1981, \mnras, 194, 809

\bibitem[\protect\citeauthoryear{{Larson}}{{Larson}}{2005}]{Larson2005}
{Larson} R.~B.,  2005, \mnras, 359, 211

\bibitem[\protect\citeauthoryear{{Larson}}{{Larson}}{2007}]{Larson2007}
{Larson} R.~B.,  2007, ArXiv Astrophysics e-prints

\bibitem[\protect\citeauthoryear{{Lubow} \& {Pringle}}{{Lubow} \&
  {Pringle}}{1993}]{LubowPringle1993}
{Lubow} S.~H.,  {Pringle} J.~E.,  1993, \mnras, 263, 701

\bibitem[\protect\citeauthoryear{{Mac Low} \& {Klessen}}{{Mac Low} \&
  {Klessen}}{2004}]{MacLowKlessen2004}
{Mac Low} M.,  {Klessen} R.~S.,  2004, Reviews of Modern Physics, 76, 125

\bibitem[\protect\citeauthoryear{{Mac Low}, {Klessen}, {Burkert} \&
  {Smith}}{{Mac Low} et~al.}{1998}]{MacLowetal1998}
{Mac Low} M.,  {Klessen} R.~S.,  {Burkert} A.,    {Smith} M.~D.,  1998,
  Physical Review Letters, 80, 2754

\bibitem[\protect\citeauthoryear{{Meyer}, {Adams}, {Hillenbrand}, {Carpenter}
  \& {Larson}}{{Meyer} et~al.}{2000}]{Meyeretal2000}
{Meyer} M.~R.,  {Adams} F.~C.,  {Hillenbrand} L.~A.,  {Carpenter} J.~M.,
  {Larson} R.~B.,  2000, Protostars and Planets IV, p.~121

\bibitem[\protect\citeauthoryear{{Monaghan}}{{Monaghan}}{1992}]{Monaghan1992}
{Monaghan} J.~J.,  1992, \araa, 30, 543

\bibitem[\protect\citeauthoryear{{Monaghan}}{{Monaghan}}{2005}]{Monaghan2005}
{Monaghan} J.~J.,  2005, Reports of Progress in Physics, 68, 1703

\bibitem[\protect\citeauthoryear{{Myers}}{{Myers}}{1983}]{Myers1983}
{Myers} P.~C.,  1983, \apj, 270, 105

\bibitem[\protect\citeauthoryear{{Myers} \& {Gammie}}{{Myers} \&
  {Gammie}}{1999}]{MyersGammie1999}
{Myers} P.~C.,  {Gammie} C.~F.,  1999, \apjl, 522, L141

\bibitem[\protect\citeauthoryear{{Pringle}, {Allen} \& {Lubow}}{{Pringle}
  et~al.}{2001}]{Pringleetal2001}
{Pringle} J.~E.,  {Allen} R.~J.,    {Lubow} S.~H.,  2001, \mnras, 327, 663

\bibitem[\protect\citeauthoryear{{Salpeter}}{{Salpeter}}{1955}]{Salpeter1955}
{Salpeter} E.~E.,  1955, \apj, 121, 161

\bibitem[\protect\citeauthoryear{{Solomon}, {Rivolo}, {Barrett} \&
  {Yahil}}{{Solomon} et~al.}{1987}]{Solomonetal1987}
{Solomon} P.~M.,  {Rivolo} A.~R.,  {Barrett} J.,    {Yahil} A.,  1987, \apj,
  319, 730

\bibitem[\protect\citeauthoryear{{V{\'a}zquez-Semadeni}, {Ballesteros-Paredes}
  \& {Klessen}}{{V{\'a}zquez-Semadeni} et~al.}{2003}]{VazquezSemadenietal2003}
{V{\'a}zquez-Semadeni} E.,  {Ballesteros-Paredes} J.,    {Klessen} R.~S.,
  2003, \apjl, 585, L131

\bibitem[\protect\citeauthoryear{{Vazquez-Semadeni}, {Passot} \&
  {Pouquet}}{{Vazquez-Semadeni} et~al.}{1995}]{VazquezSemadeni1995}
{Vazquez-Semadeni} E.,  {Passot} T.,    {Pouquet} A.,  1995, \apj, 441, 702

\bibitem[\protect\citeauthoryear{{V{\'a}zquez-Semadeni}, {Ryu}, {Passot},
  {Gonz{\'a}lez} \& {Gazol}}{{V{\'a}zquez-Semadeni}
  et~al.}{2006}]{VazquezSemadenetal2006}
{V{\'a}zquez-Semadeni} E.,  {Ryu} D.,  {Passot} T.,  {Gonz{\'a}lez} R.~F.,
  {Gazol} A.,  2006, \apj, 643, 245

\bibitem[\protect\citeauthoryear{{Williams}, {de Geus} \& {Blitz}}{{Williams}
  et~al.}{1994}]{Williamsetal1994}
{Williams} J.~P.,  {de Geus} E.~J.,    {Blitz} L.,  1994, \apj, 428, 693

\bibitem[\protect\citeauthoryear{{Wuchterl} \& {Klessen}}{{Wuchterl} \&
  {Klessen}}{2001}]{WuchterlKlessen2001}
{Wuchterl} G.,  {Klessen} R.~S.,  2001, \apjl, 560, L185

\bibitem[\protect\citeauthoryear{{Yorke} \& {Sonnhalter}}{{Yorke} \&
  {Sonnhalter}}{2002}]{YorkeSonnhalter2002}
{Yorke} H.~W.,  {Sonnhalter} C.,  2002, \apj, 569, 846

\bibitem[\protect\citeauthoryear{{Zinnecker}}{{Zinnecker}}{1982}]{Zinnecker198%
2}
{Zinnecker} H.,  1982, New York Academy Sciences Annals, 395, 226

\bibitem[\protect\citeauthoryear{{Zinnecker} \& {Yorke}}{{Zinnecker} \&
  {Yorke}}{2007}]{ZinneckerYorke2007}
{Zinnecker} H.,  {Yorke} H.~W.,  2007, \araa, 45, 481

\end{thebibliography}

\end{document}